\documentclass[preprint,12pt]{elsarticle}

\let\counterwithin\relax
\usepackage{chngcntr}
\usepackage{amsbsy}
\usepackage{bm}
\usepackage{float}
\usepackage{listings}
\usepackage{xcolor}
\usepackage{subcaption}
\usepackage{chngcntr}
\usepackage{hyperref}
\counterwithin{figure}{section}

\usepackage{amssymb}

\newcounter{bla}

\journal{Computer Physics Communications}

\begin{document}

\begin{frontmatter}

\title{On the derivatives of feed-forward neural networks}

\author[1,2]{Rabah Abdul Khalek}
\author[3,4]{Valerio Bertone}

\address[1]{Department of Physics and Astronomy, Vrije Universiteit, Amsterdam, 1081 HV, The Netherlands.}
\address[2]{Nikhef Theory Group, Science Park 105, 1098 XG Amsterdam, The Netherlands.}
\address[3]{Dipartimento di Fisica, Universit\'a di Pavia and INFN, Sezione di Pavia Via Bassi 6, I-27100 Pavia, Italy.}
\address[4]{IRFU, CEA, Universit\'e Paris-Saclay, F-91191 Gif-sur-Yvette, France.}


\begin{abstract}
  In this paper we present a \texttt{C++} implementation of the analytic
derivative of a feed-forward neural network with respect to its free
parameters for an arbitrary architecture, known as back-propagation.
We dubbed this code \texttt{NNAD} (Neural Network Analytic
Derivatives) and interfaced it with the widely-used
\texttt{ceres-solver}~\cite{ceres-solver} minimiser to fit neural
networks to pseudodata in two different least-squares problems. The
first is a direct fit of Legendre polynomials. The second is a
somewhat more involved minimisation problem where the function to be
fitted takes part in an integral. Finally, using a consistent
framework, we assess the efficiency of our analytic derivative formula
as compared to numerical and automatic differentiation as provided by
\texttt{ceres-solver}. We thus demonstrate the advantage of using
\texttt{NNAD} in problems involving both deep or shallow neural
networks.
\end{abstract}


\end{frontmatter}

\vspace{-17cm}
\begin{flushright}
Nikhef/2020-011
\end{flushright}
\vspace{17cm}

\newpage

{\bf PROGRAM SUMMARY}
\vspace{10pt}

\begin{small}
  \noindent
  {\em Program Title:} \texttt{NNAD}                                          \\
  {\em Licensing provisions}: MIT  \\
  {\em Programming language:} \texttt{C++}                                   \\
  {\em Nature of problem:} computation of the gradient of a
  feed-forward neural network with respect to its parameters (weights
  and biases). This is a proxy for the computation of the gradient of
  sensible figures of merit used in optimisation problems.\\
  {\em Solution method:} analytic derivative of a feed-forward neural
  network with arbitrary architecture by means of a recursive
  application of the chain rule.\\
\end{small}

\section{Introduction}

The extraction of information from experimental measurements often
requires parametrising some relevant quantities using suitable
functional forms that are eventually fitted to the data. In many
cases, even the general behaviour of such quantities is largely
unknown. In order to avoid unwanted parametric biases, it is
convenient to make as few assumptions as possible on the associated
functional form. This requires flexible objects able to describe a
large variety of behaviours: \textit{neural networks} (NNs) are thus
suitable candidates (see, \textit{e.g.}, Refs.~\cite{Forte:2002fg,
  Kumericki:2011rz, Ball:2017nwa} for some applications in high-energy
physics). However, the flexibility of NNs comes at the cost of a
complex multidimensional parameter space. Therefore, fitting (or
training) a NN typically requires a heavy numerical effort aimed at
finding the best fit parameters in a very complicated parameter space.

Efficient training algorithms often rely on the knowledge of the
gradient in parameter space of the figure of merit chosen to measure
the goodness of the fit. Therefore, an efficient computation of the
gradient is crucial to achieve performance. This is especially true
for complicated figures of merit involving NNs.

There are fundamentally two main approaches to the computation of a
gradient. The first is the \textit{numerical} approach. The second is
the \textit{analytic} approach that can be further branched off into:
``fully'' analytic (henceforth simply analytic) and \textit{automatic}
differentiations (See Sect.~\ref{sec:performance} for more details).

In this paper we consider feed-forward NNs with arbitrary architecture
and derive a compact back-propagation algorithm to
\textit{analytically} compute their gradient in parameter space.  We
then implement this algorithm in a \texttt{C++} code and compare it to
numerical and automatic differentiation as implemented in
\texttt{ceres-solver}~\cite{ceres-solver}.


The paper is organised as follows. In Sect.~\ref{sec:theory}, we first
define the figure of merit that we will use throughout the paper and
then derive the analytic back-propagation algorithm (\texttt{NNAD})
for a feed-forward NN. In Sect.~\ref{sec:Applications}, we use
\texttt{NNAD} in two specific cases: 1) a fit of pseudodata generated
using a Legendre polynomial and 2) a fit of functions involved in
convolution integrals. We devote Sect.~\ref{sec:performance} to assess
the performance of our algorithm as compared to numerical and
automatic differentiations. Finally, we give some conclusive remarks
in Sect.~\ref{sec:Conclusions}. \ref{app:code} provides some details
concerning the main functionalities of the \texttt{C++}
implementation.

\section{Definition and optimisation of the figure of
  merit}\label{sec:theory}

In this section we define the particular figure of merit that we will
be using throughout the paper.  The figure of merit has the purpose of
``measuring'' the degree of agreement between a set of data and some
parametric model. A common choice for the figure of merit is the
\textit{likelihood function}.

The likelihood function can be defined as the probability of observing
a given sample of data for a given set of parameters of the given
model. Let $\bm{D} = \{D_1,D_2,...D_n\}$ be a set of random variables
having a joint probability density $\mathcal{F}$ depending on a set of
parameters $\bm{\theta} = \{\theta_1, \theta_2, ..., \theta_m\}$ of
some parametric model. From the definition above, one can write the
likelihood function as:
\begin{equation}
  \mathcal{L}(\bm{\theta}| \bm{d}) = \mathcal{F}_{\bm{\theta}}(\bm{d}) \equiv \prod_i^n f_{i,\bm{\theta}}(d_i)\,,
\end{equation}
where $\bm{d}=\{d_1, d_2, ..., d_n\}$ are the observed values of
$\bm{D}$ (the data) and $f_{i,\bm{\theta}}$ is the single probability
distribution of $D_i$ given the set of parameters $\bm{\theta}$. Let
us take all $f_i$ to be mutually independent normal
distributions,\footnote{The case of dependent variables can be
  addressed by introducing the appropriate correlation matrix.}
therefore:
\begin{equation}
  \mathcal{L}(\bm{\theta}|\bm{d}) = \prod_i^n
  \frac{1}{\sqrt{2\pi\sigma_i^2}} e^{-\frac{(m_i({\bm \theta}) - d_i)^2}{2\sigma_i^2}}\,,
\end{equation} 
where $d_i$ and $\sigma_i$ are respectively the central value and the
uncertainty of the $i$-th datapoint, and $m_i$ is the corresponding
model prediction obtained with the set of parameters $\bm{\theta}$.

The goal of a regression analysis is to find a particular set of
parameters $\bm{\theta}$ that maximises the likelihood
$\mathcal{L}(\bm{\theta})$, thus maximising the probability of
observing the given data. This procedure typically goes under the name
of maximum likelihood estimation. However, finding the maximum of the
likelihood turns out to be computationally expensive. In order to
simplify the problem, one can equivalently maximise the logarithm of
the likelihood, that reads:
\begin{equation}
\ln\left(\mathcal{L}(\bm{\theta}|\bm{d})\right) = -\frac{n}{2}\ln{(2\pi)} + \sum_i^n \ln{\frac{1}{\sigma_i}} - \frac{1}{2}\sum_i^n \left(\frac{m_i(\bm{\theta})-d_i}{\sigma_i}\right)^2\,.
\end{equation} 
Since the first two terms on the right-hand side are constant and the
last term is negative, maximising the likelihood is equivalent to
minimising this last term, that is:
\begin{equation}\label{eq:MLE}
    \max_{\bm{\theta}}\ln\left(\mathcal{L}(\bm{\theta}|\bm{d})\right) \rightarrow \min_{\bm{\theta}} \sum^n_i \left(\frac{ m_i(\bm{\theta}) - d_i}{\sigma_i}  \right)^2
\equiv \min_{\bm{\theta}} \chi^2(\bm{\theta})\,.
\end{equation}
This defines the $\chi^2$ as an effective loss function to be
minimised in a regression analysis. In the studies below, we will use
the $\chi^2$ as a figure of merit.

There exist numerous strategies to solve Eq.~(\ref{eq:MLE}). In this
paper we will focus on deterministic \textit{derivative-based}
algorithms whereby the solution of Eq.~(\ref{eq:MLE}) is reached by
following the steepest descending path on the hyper-surface defined by
the $\chi^2$.  Alternatives to the deterministic approach are the
\textit{derivative-free} algorithms that broadly consist of exploring
random paths of the parameter space until the minimum of the $\chi^2$
is found.  Examples of derivative-free algorithms are the genetic
algorithm and the covariance-matrix-adaptation
evolution-strategy~\cite{1554902}.

Within deterministic derivative-based algorithms, we have chosen to
work with the quite popular Levenberg-Marquardt
algorithm~\cite{Gavin2013TheLM}. It combines the gradient-descent
algorithm to the Gauss-Newton method. When the parameters are far from
their optimal value, the Levenberg-Marquardt method behaves more like
a gradient-descent algorithm in which the $\chi^2$ is reduced by
updating the parameters along the steepest-descent direction. The
Gauss-Newton method is instead used when the parameters are close to
their optimal value to identify the minimum of the $\chi^2$ by
assuming it to be locally quadratic. Both the gradient-descent and
Gauss-Newton methods rely on the knowledge of the derivatives of the
$\chi^2$ with respect to the parameters $\bm{\theta}$. In fact, the
computation of the gradient of the $\chi^2$ is a bottleneck for
\textit{all} deterministic algorithms in general and, as such, for the
Levenberg-Marquardt one in particular. Therefore, an efficient
computation of this quantity is highly desirable.

\section{Derivative of the $\chi^2$}
\label{sec:dchi2}

In this section we discuss how to treat the gradient of the
$\chi^2$
for a model parameterised by a feed-forward NN. This will
automatically result into the derivative of the NN itself that we will
then explicitly work out.

For the sake of simplicity, we consider one single experimental point
measured at ${\pmb \xi}\in\mathbb{R}^m$ with central value
${\pmb d}\in\mathbb{R}^n$ and standard deviation
$\pmb\sigma\in\mathbb{R}^n$ that we want to fit with a NN
${\pmb N}:\mathbb{R}^m\rightarrow\mathbb{R}^n$ with $L$ layers and
parametrised by a set of weights and biases
$\{\omega_{ij}^{(\ell)},\theta_{i}^{(\ell)}\}$. The corresponding
$\chi^2$ reads:
\begin{equation}\label{eq:chi2def}
  \chi^2[\{\omega_{ij}^{(\ell)},\theta_{i}^{(\ell)}\}] =  \sum_{k=1}^n\left(\frac{ N_k({\pmb \xi};\{\omega_{ij}^{(\ell)},\theta_{i}^{(\ell)}\}) - d_k}{\sigma_k}  \right)^2\,,
\end{equation}
where $N_k$ is the $k$-th output of the NN. We also assume that all
the nodes belonging to the $\ell$-th layer have the same activation
function $\phi_\ell$. The $k$-th output of the NN can then be written
recursively as:
\begin{equation}\label{eq:nnexplicit}
\begin{array}{rcl}
  \displaystyle 
  N_k({\pmb \xi};\{\omega_{ij}^{(\ell)},\theta_{i}^{(\ell)}\}) &=&\displaystyle
                                                                   \phi_L\left(\sum_{j^{(1)}}^{N_{L-1}}\omega_{k
                                                                   j^{(1)}}^{(L)}y_{j^{(1)}}^{(L-1)}+\theta_{k}^{(L)}\right)\\
  \\
                                                               &=&\displaystyle
                                                                   \phi_L\left(\sum_{j^{(1)}=1}^{N_{L-1}}\omega_{k
                                                                   j^{(1)}}^{(L)}\phi_{L-1}\left(\sum_{j^{(2)}=1}^{N_{L-2}}\omega_{j^{(1)}
                                                                   j^{(2)}}^{(L)}y_{j^{(2)}}^{(L-2)}+\theta_{j^{(1)}}^{(L-1)}\right)+\theta_{k}^{(L)}\right)\\
                                                               &=&\dots\,.
\end{array}
\end{equation}
The nesting in Eq.~(\ref{eq:nnexplicit}) continues until the input
layer is reached. The derivative of the $\chi^2$ in
Eq.~(\ref{eq:chi2def}) with respect to the weight
$\omega_{ij}^{(\ell)}$, relevant to the computation of the gradient,
takes the form:
\begin{equation}\label{eq:chi2derdef}
  \frac{\partial\chi^2}{\partial \omega_{ij}^{(\ell)}} =  2\sum_{k=1}^n\left(\frac{ N_k({\pmb \xi};\{\omega_{ij}^{(\ell)},\theta_{i}^{(\ell)}\}) - d_k}{\sigma_k^2}  \right) \frac{\partial N_k}{\partial \omega_{ij}^{(\ell)}}\,,
\end{equation}
and similarly for the derivative with respect to the bias
$\theta_{i}^{(\ell)}$.  Eq.~(\ref{eq:chi2derdef}) reduces the
computation of the derivatives of the $\chi^2$ in
Eq.~(\ref{eq:chi2def}) to the computation of the derivatives of
${\pmb N}$. In this respect, the feed-forward structure of the NN in
Eq.~(\ref{eq:nnexplicit}) is crucial to work out an explicit
expression for such derivatives. We start by defining:
\begin{equation}
\begin{array}{l}
  \displaystyle x_i^{(\ell)} =
  \sum_{j=1}^{N_{\ell-1}}\omega_{ij}^{(\ell)}y_{j}^{(\ell-1)}+\theta_{i}^{(\ell)}\,,\\
  \\
  \displaystyle y_i^{(\ell)} =
  \phi_\ell\left(x_i^{(\ell)}\right)\,,\\
  \\
  \displaystyle z_i^{(\ell)} = \phi'_\ell\left(x_i^{(\ell)}\right)\,.
\end{array}
\end{equation}
We then apply the chain rule to compute the derivative of $N_k$
w.r.t. $\omega_{ij}^{(\ell)}$:
\begin{equation}
\begin{array}{rcl}
\displaystyle  \frac{\partial N_k}{\partial \omega_{ij}^{(\ell)}}&=&\displaystyle\frac{\partial
    y_k^{(L)}}{\partial \omega_{ij}^{(\ell)}}\\
\\
&=&\displaystyle z_k^{(L)}
  \frac{\partial x_k^{(L)}}{\partial \omega_{ij}^{(\ell)}}\\
\\
&=&\displaystyle  \sum_{j^{(1)}=1}^{N_{L-1}}  \left[z_k^{(L)}\omega_{k j^{(1)}}^{(L)}\right]\frac{\partial
    y_{j^{(1)}}^{(L-1)}}{\partial \omega_{ij}^{(\ell)}}\\
\\
&=&\displaystyle  \sum_{j^{(1)}=1}^{N_{L-1}} \sum_{j^{(2)}=1}^{N_{L-2}} \left[z_k^{(L)} \omega_{k j^{(1)}}^{(L)}
  \right]\left[z_{j^{(1)}}^{(L-1)}\omega_{j^{(1)} j^{(2)}}^{(L-1)}\right]\frac{\partial
    y_{j^{(2)}}^{(L-2)}}{\partial \omega_{ij}^{(\ell)}}\\
\\
&=&\displaystyle \dots
\end{array}
\end{equation}
As evident, the chain rule penetrates into the NN starting from the
output layer all the way back until the $\ell$-th layer (\textit{i.e.}
the layer where the parameter $\omega_{ij}^{(\ell)}$ belongs to). In
order to write the derivative in a closed form, we define the $(i,j)$
entry of the matrix $\mathbf{S}^{(\ell)}$ as:
\begin{equation}
  z_i^{(\ell)}\omega_{ij}^{(\ell)}=S_{ij}^{(\ell)} \left(= \frac{\partial y_i^{(\ell)}}{\partial
      y_j^{(\ell-1)}}\right) \,,
\end{equation}
so that:
\begin{equation}
  \frac{\partial {\pmb N}}{\partial \omega_{ij}^{(\ell)}} =
  \mathbf{S}^{(L)}\cdot \mathbf{S}^{(L-1)}\cdots
  \mathbf{S}^{(\ell+1)}\cdot\frac{\partial {\pmb y}^{(\ell)}}{\partial
    \omega_{ij}^{(\ell)}}\,,
\end{equation}
with ${\pmb y}^{(\ell)}$ the $N_\ell$-dimensional vector of all
$y_i^{(\ell)}$. This expression can be written in a more compact form
as:
\begin{equation}\label{eq:derwij}
  \frac{\partial {\pmb N}}{\partial \omega_{ij}^{(\ell)}} =
  \left[\prod_{\alpha=L}^{\ell+1}\mathbf{S}^{(\alpha)}\right]\cdot\frac{\partial {\pmb y}^{(\ell)}}{\partial
    \omega_{ij}^{(\ell)}}\,.
\end{equation}
It is important to notice that the product of matrices
$\mathbf{S}^{(\alpha)}$ in Eq.~(\ref{eq:derwij}) runs backwards from
$L$ to $\ell + 1$.  The derivative in the r.h.s. of
Eq.~(\ref{eq:derwij}) can finally be computed explicitly and for the
$k$-th component of ${\pmb y}^{(\ell)}$ it reads:
\begin{equation}\label{eq:dydom}
\frac{\partial y_k^{(\ell)}}{\partial
\omega_{ij}^{(\ell)}} = z_k^{(\ell)} \frac{\partial x_k^{(\ell)}}{\partial
\omega_{ij}^{(\ell)}} = \delta_{ki} z_i^{(\ell)}y_j^{(\ell-1)}\,.
\end{equation}
Similarly, the derivative of ${\pmb N}$ w.r.t. the bias $\theta_{i}^{(\ell)}$ takes the form:
\begin{equation}
\frac{\partial {\pmb N}}{\partial \theta_{i}^{(\ell)}} =
\left[\prod_{\alpha=L}^{\ell+1}\mathbf{S}^{(\alpha)}\right]\cdot\frac{\partial {\pmb y}^{(\ell)}}{\partial
\theta_{i}^{(\ell)}}\,,
\end{equation}
with:
\begin{equation}\label{eq:dydth}
\frac{\partial y_k^{(\ell)}}{\partial
\theta_{i}^{(\ell)}} = \delta_{ki}z_i^{(\ell)}\,.
\end{equation}
The presence of $\delta_{ki}$ in both Eqs.~(\ref{eq:dydom}) and~(\ref{eq:dydth}) simplifies the
computation yielding:
\begin{equation}\label{eq:derivativesfinal}
\begin{array}{rcl}
  \displaystyle\frac{\partial N_k}{\partial \theta_{i}^{(\ell)}} &=&\displaystyle
                                                                     {\Sigma}_{ki}^{(\ell)}
                                                                     z_i^{(\ell)}\,,\\
  \\
  \displaystyle\frac{\partial N_k}{\partial \omega_{ij}^{(\ell)}} &=&\displaystyle{\Sigma}_{ki}^{(\ell)}
                                                                      z_i^{(\ell)} y_j^{(\ell-1)}\,,
\end{array}
\end{equation}
where ${\Sigma}_{ki}^{(\ell)}$ is the $(k,i)$ entry of the matrix:
\begin{equation}
\mathbf{\Sigma}^{(\ell)} = \prod_{\alpha=L}^{\ell+1}\mathbf{S}^{(\alpha)}\,.
\end{equation}
The identities in Eq.~(\ref{eq:derivativesfinal}) can finally be used
to compute the gradient of the $\chi^2$ through
Eq.~(\ref{eq:chi2derdef}) (and its respective for the bias
$\theta_{i}^{(\ell)}$).  From the point of view of a numerical
implementation, it is crucial to notice that the matrix
$\mathbf{\Sigma}^{(\ell)}$ can be computed recursively moving
backwards (hence the name back-propagation) from the output layer as:
\begin{equation}
\mathbf{\Sigma}^{(\ell-1)} = \mathbf{\Sigma}^{(\ell)} \cdot \mathbf{S}^{(\ell)}\,,
\end{equation}
starting from the initial condition:
\begin{equation}
\mathbf{\Sigma}^{(L)} = \mathbf{I}\,.
\end{equation}
This feature allows one to compute the derivatives w.r.t. all free
parameters of a NN with a \textit{single} iteration of the chain
rule. We point out that the iterative nature of
Eq.~(\ref{eq:derivativesfinal}) is a direct consequence of the
structure of the object being derived, \textit{i.e.} a feed-forward
NN. Therefore, Eq.~(\ref{eq:derivativesfinal}) does \textit{not}
generally apply to any artificial NN.

The implementation of Eq.~(\ref{eq:derivativesfinal}) is made public
through a \texttt{C++} code that we dubbed \texttt{NNAD}, for
Neural-Network Analytic Derivatives. This code provides an interface
to construct and compute feed-forward NNs with arbitrary architecture
along with their derivarives w.r.t. weights and biases. A brief
discussion on the main features and usage of this code can be found in
\ref{app:code}.

\section{Applications}\label{sec:Applications}

In this section, we present two applications of \texttt{NNAD} where we
implemented Eq.~(\ref{eq:derivativesfinal}) derived in the previous
section. The first application is a fit of a NN to pseudodata
generated using an oscillating Legendre polynomial as an underlying
law. This represents the simplest environment in which the NN needs to
adapt locally to the single data points. The second example is instead
more involved as the NN to be fitted appears inside a convolution
integral. As a consequence, each single data point has a non-local
impact on the NN, making the fit harder.  This example is reminiscent
of fits of collinear parton-distribution and fragmentation functions
but with some simplification aimed at highlighting the robustness of
our implementation.

In order to carry out the studies outlined above, we have developed a
separate code, named \texttt{NNAD-Interface}, meant to interface
\texttt{NNAD} to other publicly available tools. Specifically,
\texttt{NNAD-Interface} is interfaced to the \texttt{ceres-solver}
minimiser~\cite{ceres-solver}, an open source \texttt{C++} library for
modeling and solving large optimisation problems that has been used in
production at Google since 2010, and to
\texttt{APFELgrid}~\cite{Bertone:2016lga} that provides a fast
computation of the convolution integrals.

\subsection{Fitting Legendre polynomials}\label{subsec:legendre}

As a first application, we take the simple example of fitting
pseudodata generated using a Legendre polynomial as an underlying
law. To this purpose, we generated $N_{\rm data}=100$ pseudodata
points over an equally-spaced grid $\{\xi_i\}$, with
$i=1,\dots,N_{\rm data}$ and $\xi_i \in [-1,1]$. The corresponding
sets of central values $\{d_i\}$ and uncertainties $\{\sigma_i\}$ is
obtained as:
\begin{equation}\label{eq:legendrepoints}
\begin{array}{l}
    d_i = [1 + P_{10}(\xi_i)]\times\mathcal{G}(1,0.1)\,,\\
\\
  \sigma_i = [1+ P_{10}(\xi_i)]\times\mathcal{G}(0,0.1)\,,
\end{array}
\end{equation}
where $P_{n}$ is the Legendre polynomial of degree $n$ and
$\mathcal{G}(\mu,\sigma)$ is the normal distribution with mean value
$\mu$ and standard deviation $\sigma$. The shift by $1$ in both
equations in Eq.~(\ref{eq:legendrepoints}) has the goal to make the
underlying law positive definite and facilitate thus the generation of
the pseudodata.

The model used to fit the data is a NN with one input node, one output
node, and a single hidden layer with 25 fully-connected nodes
(architecture $[1,25,1]$) for a total of 76 free parameters.

\begin{figure}[t]
\centering
\includegraphics[width=0.8\textwidth]{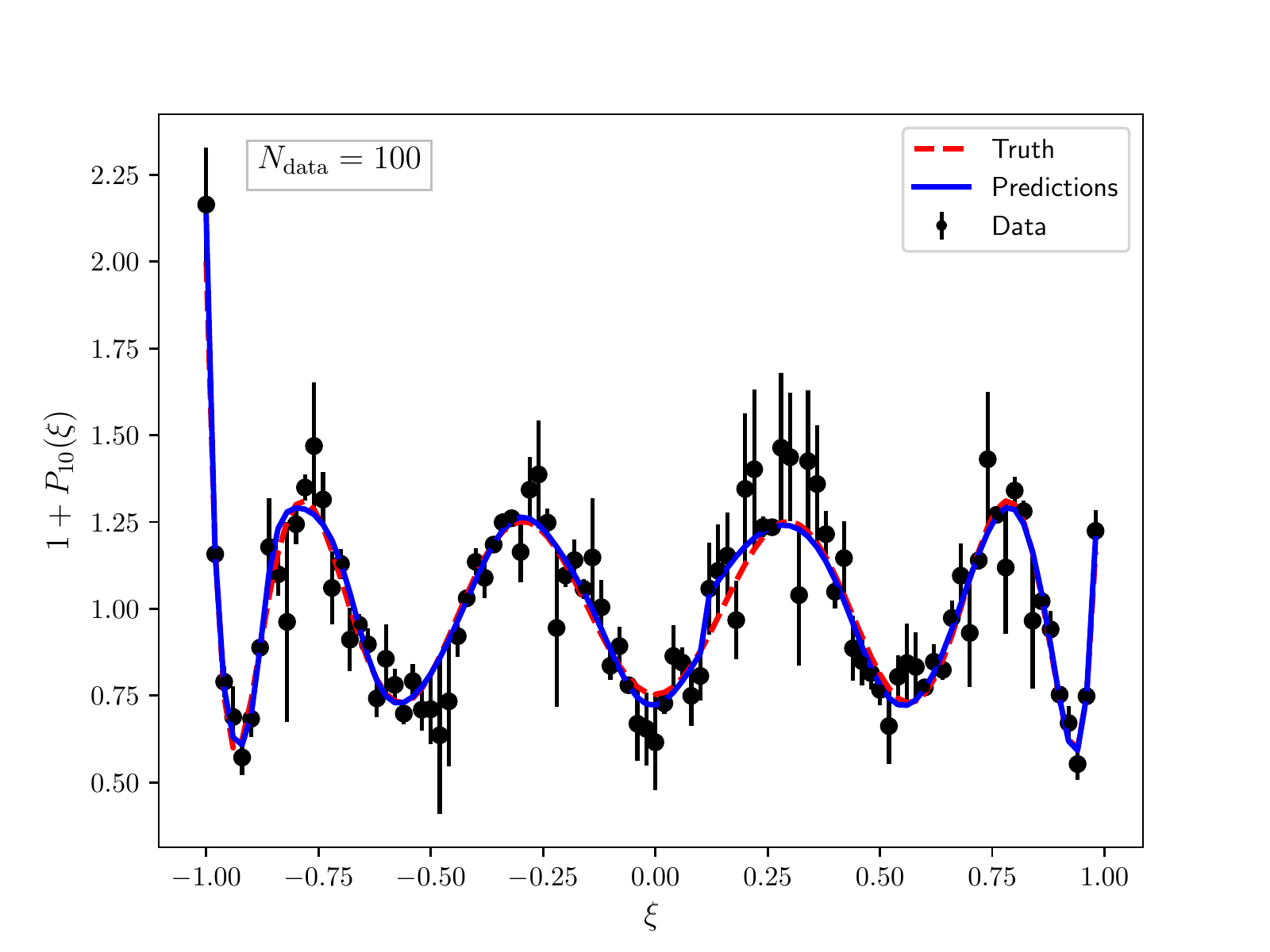}
\caption{\label{fig:P10_fit} Fit of a NN with architecture $[1,25,1]$
  to 100 pseudodata points generated according to
  Eq.~(\ref{eq:legendrepoints}). The black points represent the
  pseudodata with the corresponding uncertainty, the red line is
  underlying law (truth), and the blue line is the result of a
  1000-iteration fit.}
\end{figure}

Fig.~\ref{fig:P10_fit} shows the result of a fit of 1000 iterations.
It is evident that the fitted NN reproduces the underlying law quite
accurately. This provides a first proof of the correctness of our
derivation, Eq.~(\ref{eq:derivativesfinal}).

\subsection{Fitting parton-distribution functions}
\label{subsec:pdfs}

As a further application, we consider a somewhat more complicated but
realistic problem in which the functions to be determined appear
inside a convolution integral. This makes the fit harder in that each
single data point has a delocalised impact on the functions being
fitted. Purposely, this procedure resembles very closely the
extraction of collinear parton-distribution functions
(PDFs)~\cite{Ball:2017nwa, AbdulKhalek:2019mzd} or fragmentation
functions~\cite{Bertone:2017tyb} from experimental data. However, we
stripped the problem of some complications unnecessary to the purpose
of showing the robustness of our analytic back-propagation
implementation.

First, we consider uncorrelated data points. This makes the
computation of the $\chi^2$ easier due to the absence of off-diagonal
terms in the covariance matrix.\footnote{As stated above, accounting
  for correlations would simply amount to introducing a covariance
  matrix in the definition of the $\chi^2$.} Second, we consider an
unphysical case in which theoretical predictions only depend on two
independent unknown functions.\footnote{In the simplest case of
  inclusive deep-inelastic-scattering cross sections, the number of
  independent function is three.} With this setup, the $\chi^2$ takes
the form:
\begin{equation}\label{eq:chi2conv}
  \chi^2[\{\omega_{ij}^{(\ell)},\theta_{i}^{(\ell)}\}] =
  \sum_{k=1}^{N_{\rm data}}\left(\frac{\hat{F}(\xi_k,\zeta_k;\{\omega_{ij}^{(\ell)},\theta_{i}^{(\ell)}\}) - d_k}{\sigma_k}  \right)^2\,.
\end{equation}
The theoretical predictions $\hat{F}$ are computed in terms of
convolution integrals w.r.t. the variable $\xi$ between two distinct
sets of quantities: $C_i$ and $N_i$, with $i=1,2$:
\begin{equation}\label{eq:strctfunc}
\hat{F}(\xi,\zeta)\equiv \sum_{i=1}^2 \left[C_i \otimes_\xi N_i\right](\xi,\zeta) \equiv \left[{\pmb C}\otimes_\xi{\pmb N}\right](\xi,\zeta)\,,
\end{equation}
where $C_i$ are know functions of $\xi$ and $\zeta$,\footnote{In our
  case, $C_1$ and $C_2$ are related to the short-distance cross
  sections of the deep-inelastic-scattering process computed to
  next-to-leading order accuracy in perturbative Quantum
  Chromodynamics (QCD) and provided by the \texttt{APFEL}
  library~\cite{Bertone:2013vaa}. The variables $\xi$ and $\zeta$ can
  thus be respectively identified with the partonic momentum fraction
  $x$ and the negative squared invariant mass $Q^2$ of the vector
  boson.} while $N_i$ only depend on $\xi$ and are the functions to be
extracted. The convolution sign $\otimes_\xi$ is defined as:
\begin{equation}
\left[f\otimes_\xi g\right](\xi) \equiv \int_0^1dy\int_0^1dz\,f(y)g(z)\delta(yz-\xi)\,.
\end{equation}
In order to compute these convolution integrals very efficiently, we
employ the \texttt{APFELgrid} fast interface~\cite{Bertone:2016lga}.

Finally, the computation of the gradient of the $\chi^2$ in
Eq.~(\ref{eq:chi2conv}) requires the following derivatives:
\begin{equation}\label{eq:deromegatheta}
\begin{array}{rcl}
  \displaystyle \frac{\partial \chi^2}{\partial \omega_{ij}^{(\ell)}} &=&\displaystyle
                                                                          2\sum_{k=1}^{N_{\rm data}}\left(\frac{\left[{\pmb C}\otimes_\xi{\pmb
                                                                          N}\right](\xi_k,\zeta_k)-d_k}{\sigma_k^2}\right)\left[ {\pmb C}\otimes_\xi \frac{\partial {\pmb N}}{\partial \omega_{ij}^{(\ell)}}\right](\xi_k,\zeta_k)\,,\\
  \\
  \displaystyle\frac{\partial \chi^2}{\partial \theta_{i}^{(\ell)}} &=&\displaystyle
                                                                        2\sum_{k=1}^{N_{\rm data}}\left(\frac{\left[{\pmb C}\otimes_\xi{\pmb
                                                                        N}\right](\xi_k,\zeta_k)-d_k}{\sigma_k^2}\right)\left[
                                                                        {\pmb
                                                                        C}\otimes_\xi
                                                                        \frac{\partial
                                                                        {\pmb
                                                                        N}}{\partial
                                                                        \theta_{i}^{(\ell)}}\right](\xi_k,\zeta_k)\,,
\end{array}
\end{equation}
involving derivatives of the NN that we compute with \texttt{NNAD}.

The central values of the pseudodata $\{d_i\}$ are generated
convoluting the functions $C_1$ and $C_2$ with the appropriate
combinations of PDFs taken from the
\texttt{NNPDF31\_nlo\_pch\_as\_0118} set of the NNPDF3.1
family~\cite{Ball:2017nwa} accessed through the \texttt{LHAPDF}
interface~\cite{Buckley:2014ana}. The set of variables
$(\xi_k,\zeta_k)$ is such that, for each of the five values of
$\zeta_k\in \{3.5,\,22.5,\,144.9,\,932.5,\,6000\}$, 200 points
logarithmically distributed in the interval $[10^{-3},\,1]$ are
generated for the variable $\xi_k$. This amounts to a total of
$N_{\rm data}=1000$ pseudodata points. In addition, for each point
$d_i$ an uncertainty $\sigma_i$ is randomly generated according to:
\begin{equation}
  \sigma_i = d_i\times\mathcal{U}(0.05,\,0.07)\,,
\end{equation}
where $\mathcal{U}(a,\,b)$ is a uniform distribution between $a$ and
$b$ and zero elsewhere. In other words, the pseudodata points have a
relative uncertainty between 5\% and 7\%. Fig.~\ref{fig:F2p} displays
the full set of pseudodata points used in our fit as a function of
$\xi$ for the different values of $\zeta$.
\begin{figure}[t]
  \centering
  \includegraphics[width=0.8\textwidth,keepaspectratio]{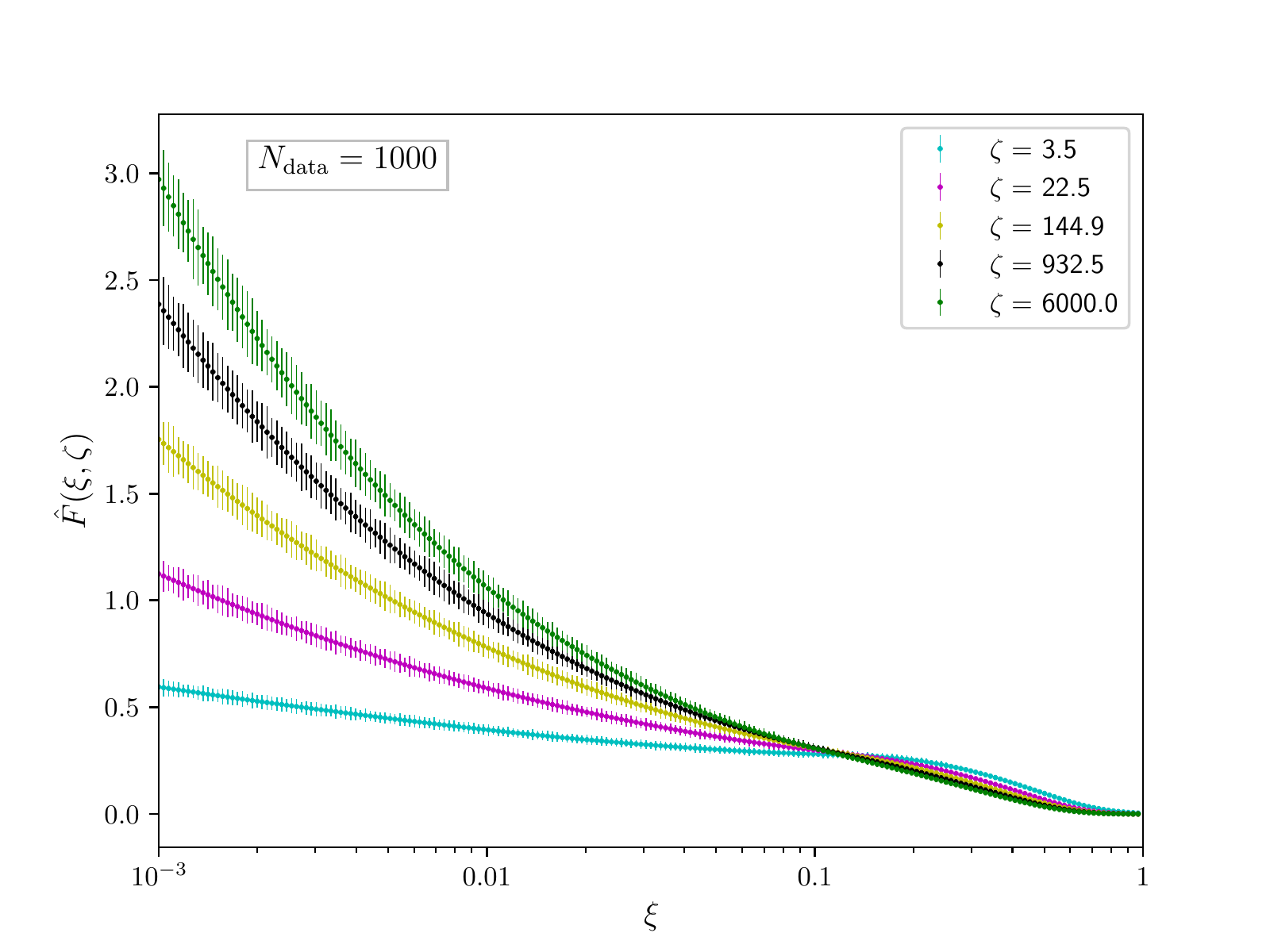}
  \caption{\label{fig:F2p} The pseudodata set for the observable
    $\hat{F}(\xi,\zeta)$, Eq.~(\ref{eq:strctfunc}), used in our
    determination of the functions $N_1$ and $N_2$. The data points
    correspond to 200 points in $\xi$ logarithmically distributed in
    the interval $[10^{-3},\,1]$ for each of 5 different values of
    $\zeta$. The uncertainties are uniformly distributed between $5\%$
    and $7\%$.}
\end{figure}

For ${\pmb N}$ in Eq.~(\ref{eq:strctfunc}) we have chosen the
architecture $[2,10,2]$\footnote{While the two output nodes correspond
  to $N_1$ and $N_2$, the two input nodes correspond to $\xi$ and
  $\ln(\xi)$. This particular input configuration seems to help the
  convergence of the fit.} for a total of 52 free parameters. The
complexity of such a large parameter space may lead to a possible
dependence of the results on the starting point of the fit in the
parameter space. To avoid this problem, we performed $N_{\rm rep}=100$
replica fits of 1000 iterations each, initialising the parameters of
the NN randomly in the interval $[-1,1]$ with a uniform
distribution. The results will then be presented as averages over the
100 replicas with uncertainties given by the standard deviation.

Fig.~\ref{fig:F2p} shows the predictions obtained after the fit,
normalised to the central value of the pseudodata, plotted as a
function of $\xi$ for all values of $\zeta$. The vertical error bars
correspond to the pseudodata uncertainties. The agreement is evidently
excellent. Indeed, the predictions overlap perfectly with the central
value of the pseudodata. In addition, the uncertainty band of the
predictions (plotted but not visible) is much smaller than the
pseudodata uncertainties. This last observation leads to the
conclusion that the final results are almost completely insensitive to
the starting point of the fit in the parameter space.
\begin{figure}[t]
  \centering
  \includegraphics[width=0.8\textwidth,keepaspectratio]{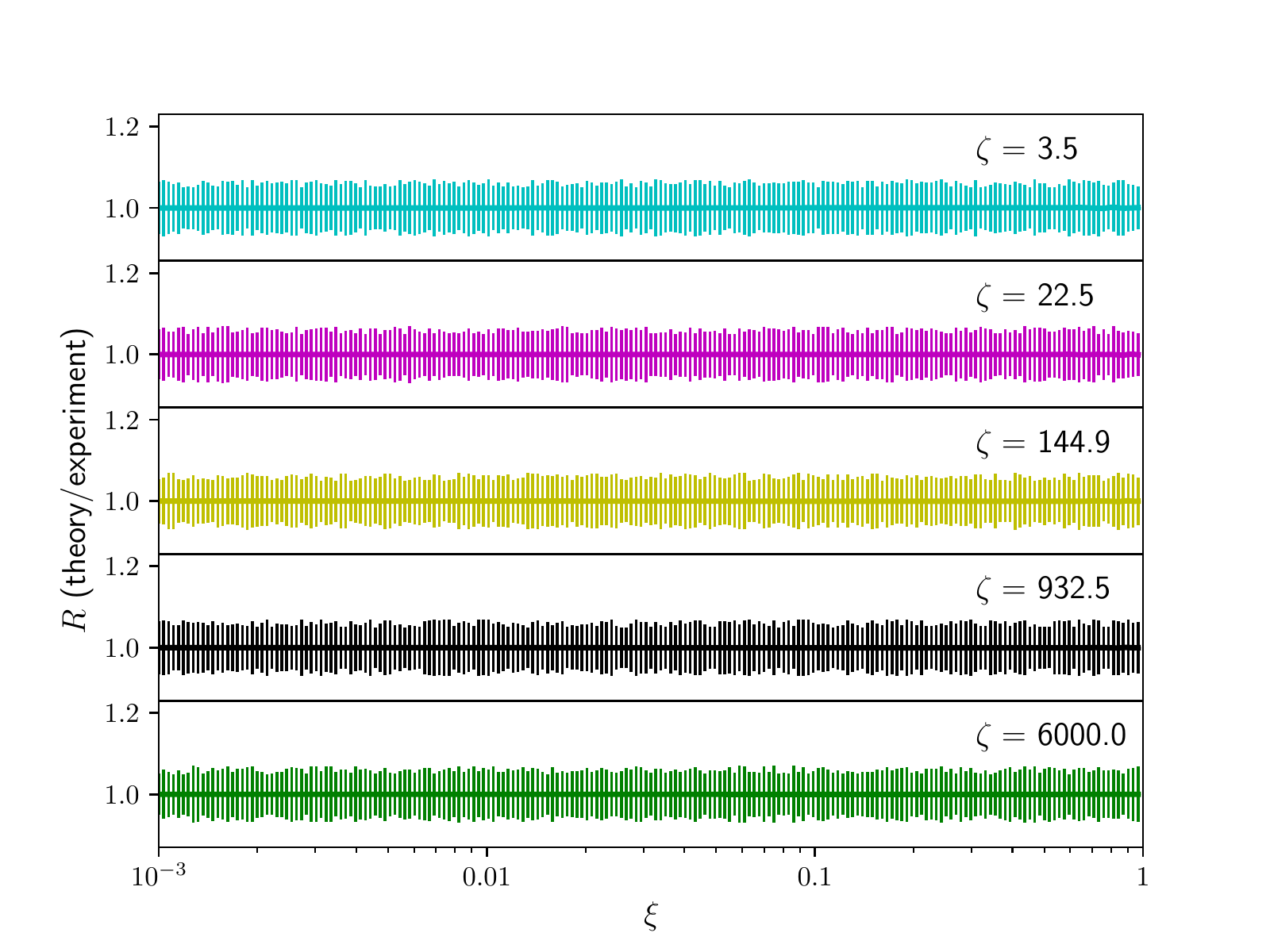}
  \caption{\label{fig:F2p} Ratio of predictions over data for the
    observable $\hat{F}(\xi,\zeta)$ for all the data included in the
    fit. The ratios are plotted as functions of $\xi$ for all the
    different values of $\zeta$.}
\end{figure}

\begin{figure}
  \centering
  \includegraphics[width=0.8\textwidth,keepaspectratio]{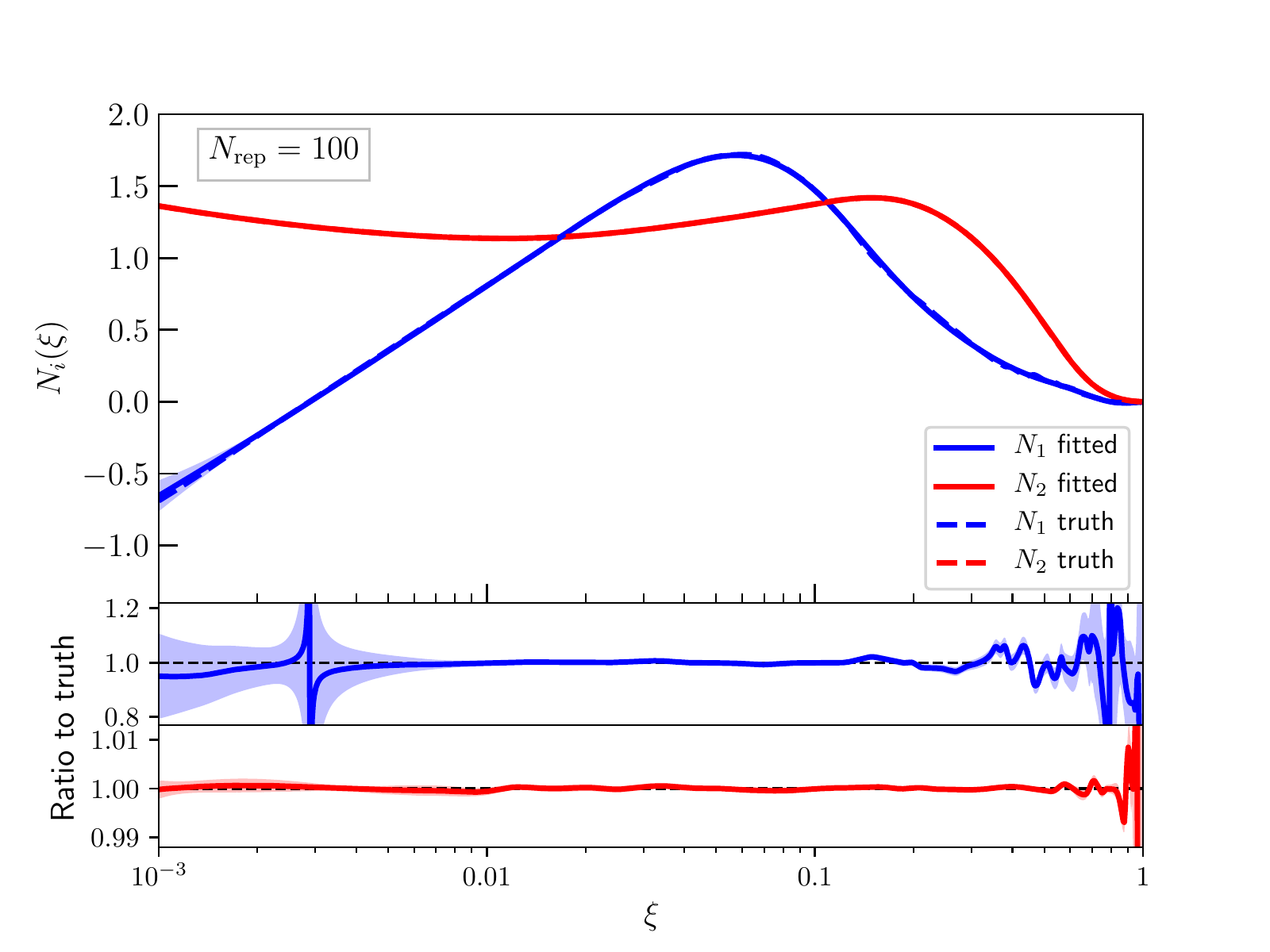}
  \caption{\label{fig:PDFs} The central value and standard deviation
    of the fitted $N_1$ and $N_2$ compared to their respective
    ``truth'' taken from \texttt{NNPDF31\_nlo\_pch\_as\_0118}.}
\end{figure}

Finally, it is interesting to compare the fitted quantities $N_1$ and
$N_2$ appearing in Eq.~(\ref{eq:strctfunc}) to their ``truth'',
\textit{i.e.} the functions used to produce the pseudodata. This
comparison is shown in Fig.~\ref{fig:PDFs}. As expected, the agreement
is excellent, particularly for the function $N_2$.\footnote{This was
  to be expected because, by construction, the observable $\hat{F}$ is
  more sensitive to the function $N_2$ than to $N_1$. The reason is
  that, in the perturbative expansion of $\hat{F}$ in terms of the QCD
  coupling, $N_2$ contributes starting from the leading order while
  $N_1$ enters at the next order. Therefore, the contribution of $N_1$
  to $\hat{F}$ is numerically less significant than that of $N_2$.}
The lower insets of Fig.~\ref{fig:PDFs} clearly show that the spread
deriving from starting the fit from different points in the parameter
space is generally small. This points to the fact that, despite the
complexity of the parameter space, all the fits converge towards the
same minimum of the $\chi^2$ irrespective of the starting point.
In conclusion, the ability of fitting functions involved in
convolution integrals is a further validation of the goodness of our
derivative formula in Eq.~(\ref{eq:derivativesfinal}).

\section{Performance assessment}\label{sec:performance}

In this section, we gauge the performance advantage of using the
\textit{analytic} back-propagation formula
Eq.~(\ref{eq:derivativesfinal}), detailed in Sect.~\ref{sec:dchi2} and
used in Sect.~\ref{sec:Applications}, over the \textit{automatic} and
\textit{numerical} methods as implemented in
\texttt{ceres-solver}~\cite{ceres-solver}.

Numerical differentiation only relies on the knowledge of the function
to be differentiated in the vicinity of the differentiation point, in
a way that derivatives are computed in terms of incremental
ratios. This method has the advantage that it applies
straightforwardly to any function.  The drawback is that it is
typically slower and potentially less accurate than an analytic
approach.

Automatic differentiation is instead an analytic approach that relies
on dual numbers (or \texttt{Jets} in the language of
\texttt{ceres-solver}).  In our context, a \texttt{Jet} represents a
parameter and its derivatives w.r.t. all other parameters in the
computational graph that either precede it (in forward-mode) or
succeed it (in backward-mode). Specifically, it contains a scalar
value corresponding to the parameter itself and the vector of its
derivatives. With a set of elementary operations, it allows for an
iterative application of the derivative chain rule so that any
parameter derivative could be easily computed from another connected
one in the graph.  Our method closely follows this procedure. However,
as pointed out in Sect.~\ref{sec:dchi2},
Eq.~(\ref{eq:derivativesfinal}) has the advantage that the full set of
derivatives is obtained in one shot iterating backwards through the
NN. Conversely, standard automatic derivatives are typically computed
in batches - not simultaneously - and not necessarily taking advantage
of any possible iterative structure of the function being
differentiated.  Nonetheless, it should be pointed out that automatic
differentiation is a totally general method, not specific to
feed-forward NNs as our method is.

In order to compare the efficiency of the different differentiation
strategies, we considered the same minimisation problem discussed in
Sect.~\ref{subsec:legendre} in which a NN with a single hidden layer
was fitted to pseudodata produced using a (shifted) Legendre
polynomial of degree 10 as an underlying law. As in
Sect.~\ref{subsec:legendre}, each fit was stopped at the 1000-th
iteration but the data set was extended from $N_{\rm data} = 100$ to
$N_{\rm data} = 1000$ pseudodata points evenly distributed in the
interval $[-1,1]$.

\begin{figure}[t]
\centering
\includegraphics[width=0.8\textwidth]{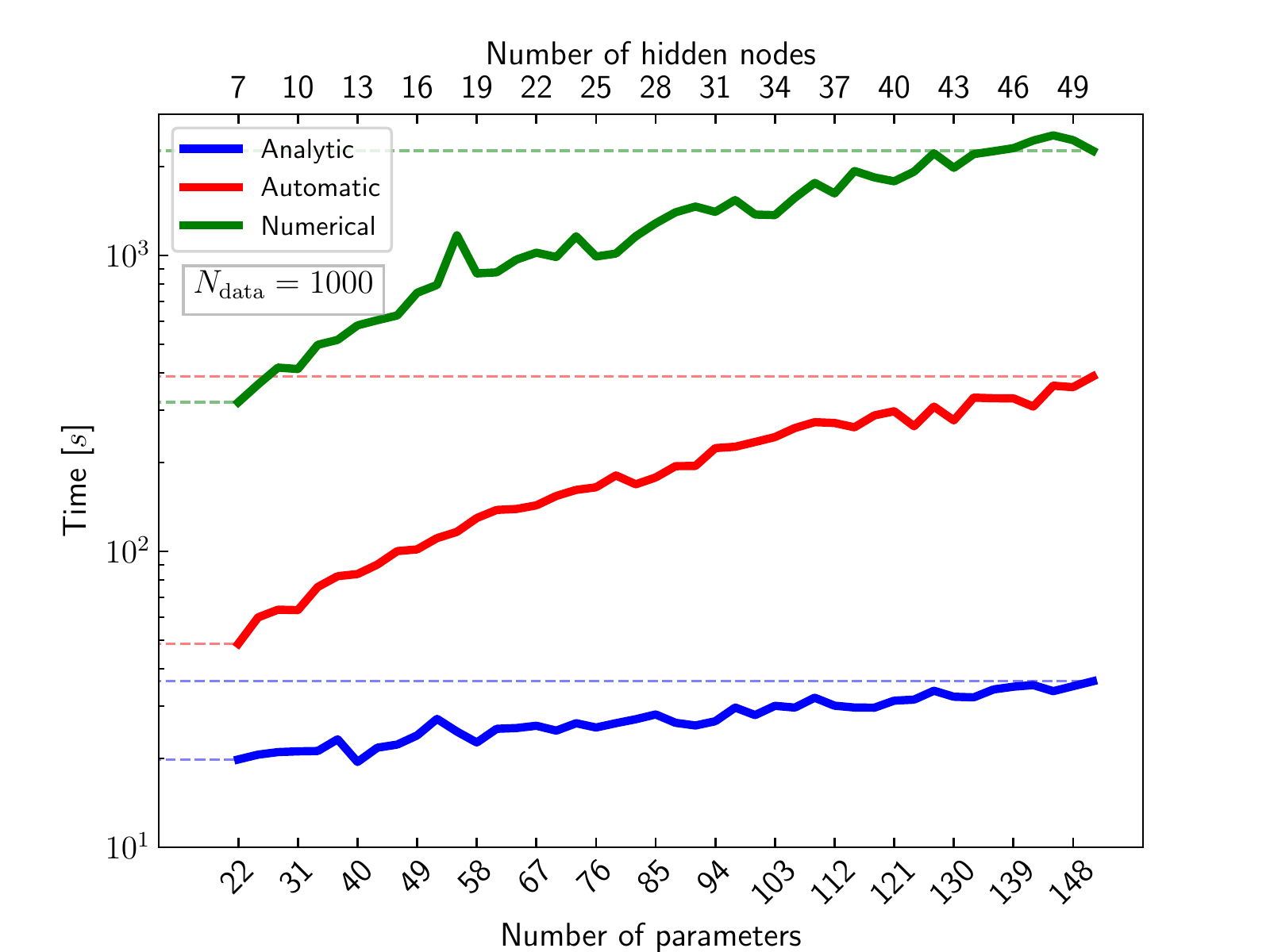}
\caption{\label{fig:Perf} Comparison of the performance of the three
  differentiation strategies. The plot shows the minimisation time as
  a function of the number of parameters/nodes for fits of 1000
  iterations of a single-layer NN to a Legendre polynomial of degree
  10.}
\end{figure}

The comparison is done by recording, for all the three differentiation
strategies, the fitting time as a function of the number of nodes of
the hidden layer.\footnote{For a single-layer NN, the number of
  parameters increases linearly with the number of nodes.}  Despite
the fitting time for each differentiation strategy only depends on the
number of iterations (1000 in our case), we required that only
successful fits\footnote{We define a successful fit to be one with
  $\chi^2/N_{\rm data} < 1$.} that converged towards the global
minimum of the $\chi^2$ were considered. To this end, for every
architecture, we performed 50 independent fits to the same data set
with the NN initialised differently at each fit. We then selected the
running time of that with the smallest $\chi^2$. The result of this
procedure for the three differentiation strategies is shown in
Fig.~\ref{fig:Perf}.  Somewhat expectedly, the numerical
differentiation (green curve) is by far the least efficient (note the
logarithmic scale on the $y$ axis), being around a factor 5 slower
than the automatic differentiation (red curve) and by a factor between
15 (for 22 parameters) and 50 (for 148 parameters) slower than the
analytic differentiation (blue curve). More surprising is instead the
difference between automatic and analytic differentiations. While the
departure is smaller for relatively small numbers of parameters, it
steadily increases as the number of parameters grows, with automatic
differentiation being around 10 times slower than the analytic one for
148 parameters. This behaviour suggests that the iterative structure
of Eq.~(\ref{eq:derivativesfinal}) provides a performance advantage
w.r.t. standard automatic differentiation as implemented in
\texttt{ceres-solver}.\footnote{Possibly, part of the degradation in
  performance of the automatic differentiation stems from the fact
  that the figure of merit gets evaluated multiple times, computing a
  small set of derivatives at each pass. The size of this set is
  controlled by \texttt{kStride} in \texttt{ceres-solver} that we take
  to be equal to 1 in all our tests.}

In order to support this hypothesis, we have increased the number of
hidden layers of the NN (depth), this time limiting the comparison to
automatic and analytic derivatives only. The conjecture is that deeper
NNs make automatic differentiation increasingly more expensive, the
reason being that the chain of derivatives to be computed is on
average longer for a NN with more layers.  Despite this should be the
case also for the analytic back-propagation,
Eq.~(\ref{eq:derivativesfinal}), the performance worsening is expected
to be milder. The reason is that the derivative w.r.t. a parameter
belonging to a given layer only depends on the derivatives of the
parameters in the preceding layer. Since at each iteration of a fit
the derivatives w.r.t.  \textit{all} free parameters are used, our
approach is advantageous in that a single iteration through the NN
allows one to obtain them all.
\begin{figure}[t]
  \hspace{-8pt}\includegraphics[width=0.56\textwidth]{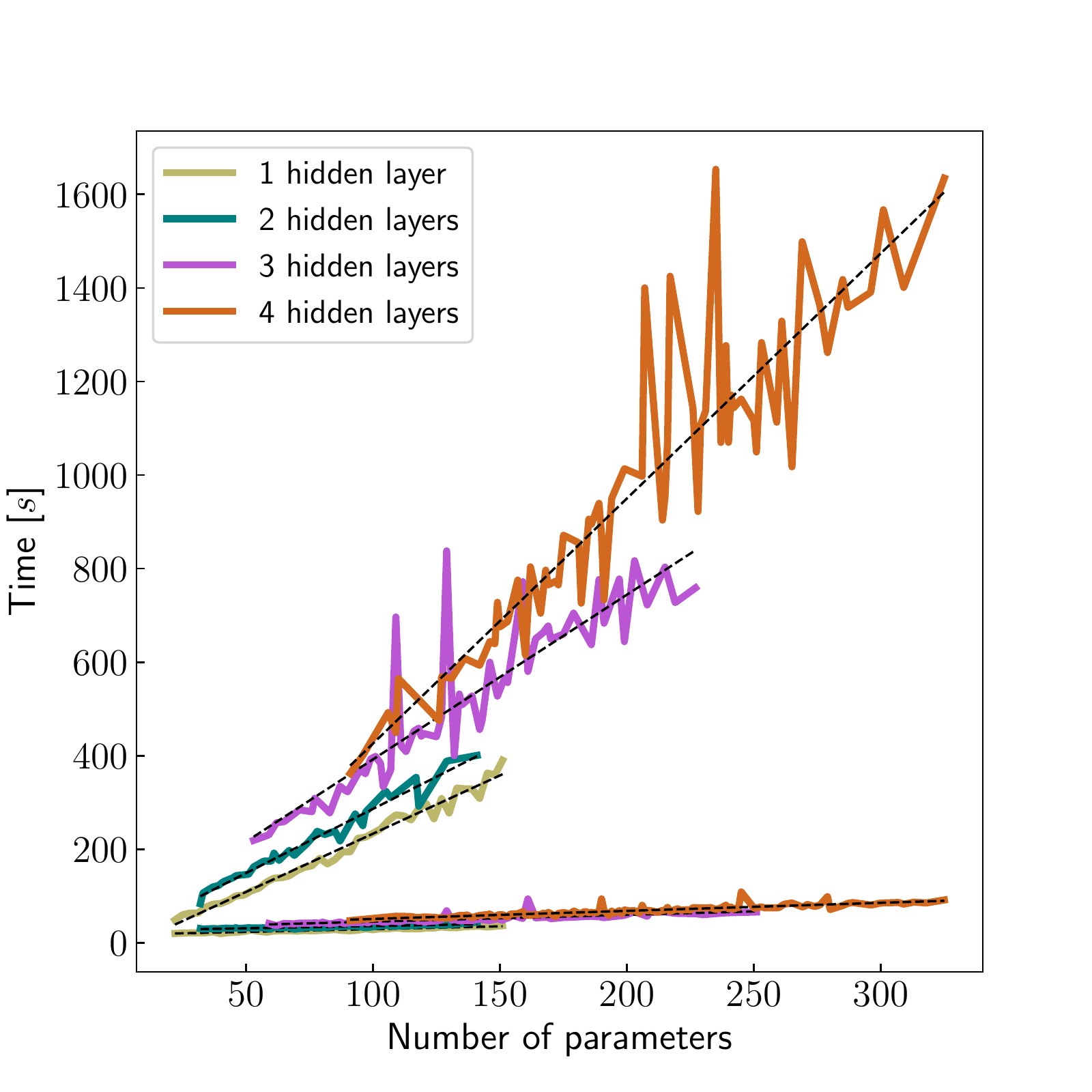}\hspace{-20pt}
  \includegraphics[width=0.56\textwidth]{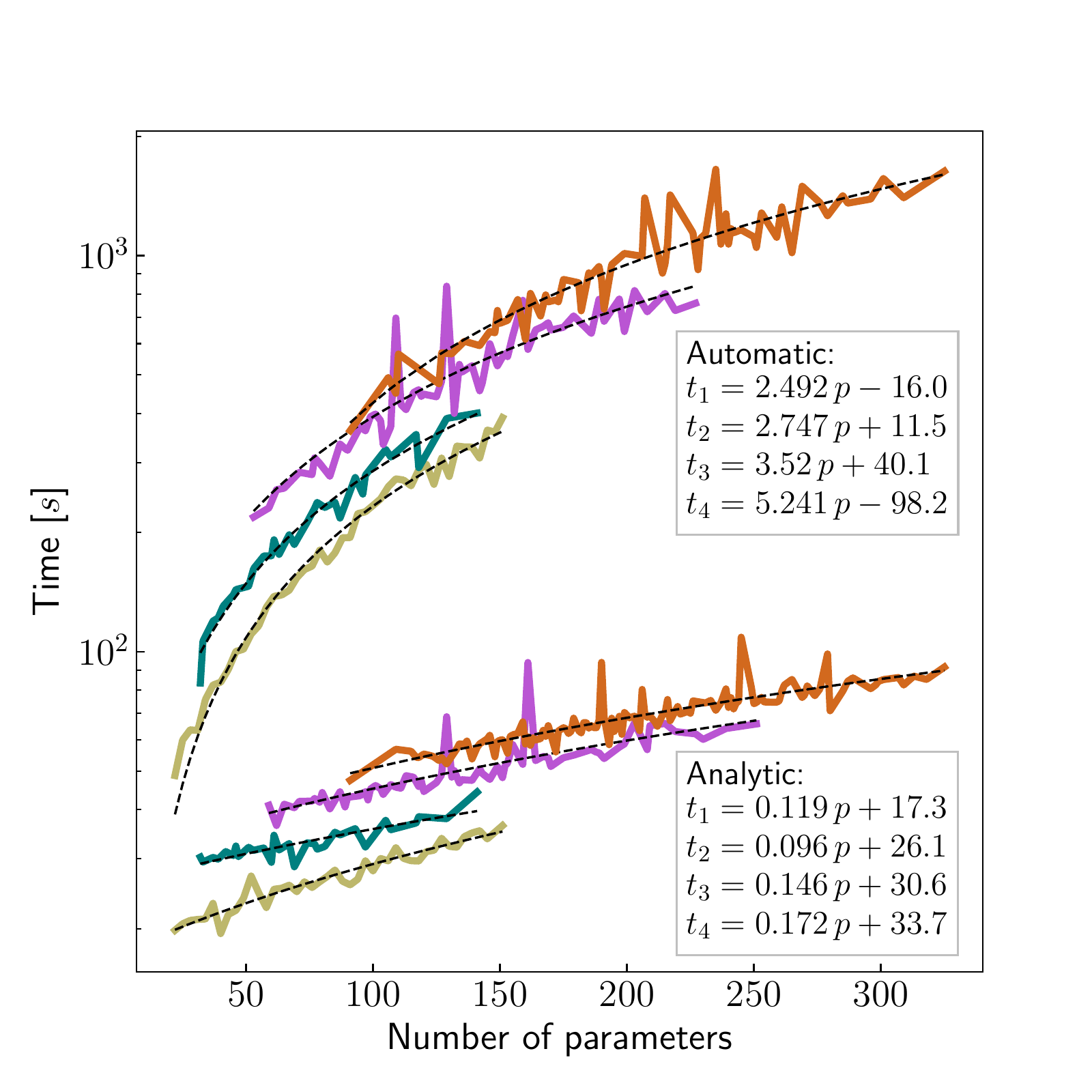}
  \caption{\label{fig:times} Performance comparison between automatic
    (\texttt{ceres-solver}) and analytic (\texttt{NNAD})
    differentiations based on the minimisation problem discussed in
    Sect.~\ref{subsec:legendre}. Different NN depths are also probed
    to highlight the differences between the two methods.}
\end{figure}
Fig.~\ref{fig:times} clearly shows a different pattern between the
analytic curves (lower part of the plot) and the automatic ones (upper
part) as the depth increases. The left and right panels display the
same curves in linear and logarithmic scale on the $y$ axis,
respectively: the latter has the goal to make the single curves
distinguishable, while the former highlights the difference in slope.
To make the comparison more quantitative, we have fitted the curves
with a straight line and reported slope and offset in the right panel
of Fig.~\ref{fig:times}. We observe the following differences:
\begin{itemize}
\item In the analytic case, the performance proportionality with the
  number of parameters remains roughly constant across all NN
  depths. This is proven by the limited variation of the slopes and
  particularly evident in the left panel of Fig.~\ref{fig:times} where
  the corresponding curves are remarkably flat.

\item Conversely, in the automatic case, the slope increases
  significantly by increasing the NN depth.  This is also evident from
  the left panel of Fig.~\ref{fig:times} and the numerical values of
  the slopes.
\end{itemize}
From the right panel of Fig.~\ref{fig:times}, it is easier to see that
both analytic and automatic curves tend to shift upwards as the NN
depth is increased.\footnote{The automatic curve with 4 hidden layers
  seems not to follow this trend for small numbers of
  parameters. However, this curve is particularly noisy and the
  straight-line offset may not be a reliable estimate of the vertical
  shift.} This is a common feature probably because the time spent to
iterate over the layers of a deeper NN is longer in both cases.

The observations concerning the performance of the different
differentiation strategies seem to support our conjecture that the
iterative structure of Eq.~(\ref{eq:derivativesfinal}) is more
efficient than numerical and standard automatic differentiations as
implemented in \texttt{ceres-solver}. More specifically, our analytic
derivation is accomplished through a reduced amount of computations,
making it particularly efficient for large-scale problems involving
deep feed-forward NNs.

\section{Conclusions}
\label{sec:Conclusions}

In this paper, we derived a compact expression for the analytic
derivatives of a feed-forward NN w.r.t. its free parameters known as
back-propagation (Eq.~(\ref{eq:derivativesfinal})) and implemented it
in the \texttt{C++} \texttt{NNAD} code (see
\href{https://github.com/rabah-khalek/NNAD}{NNAD.git}).

We presented two applications: a fit of pseudodata generated using a
Legendre polynomial and a fit of functions involved in convolution
integrals. These applications, implemented in the
\texttt{NNAD-Interface} repository (see
\href{https://github.com/rabah-khalek/NNAD-Interface}{NNAD-Interface.git}),
prove the robustness of our analytic derivation.

We finally assessed the performance gain of using our analytic
derivatives over automatic and numeric differentiation as provided by
\texttt{ceres-solver}~\cite{ceres-solver}. In order to do so, we
measured the time spent to fit the ``Legendre'' pseudodata as a
function of the number of free parameters of the NN for each of the
three methods. On top of the expected performance improvement
w.r.t. to numeric differentiation, we find that our analytic
implementation is also significantly faster than the automatic one
provided by \texttt{ceres-solver}. We also find that the improvement
does not only trivially scale with the number of parameters, but also
with the depth of the NN architecture. We interpreted this behaviour
as a consequence of the exploitation of the specific recursive
structure of a feed-forward NN in our analytic back-propagation
formula. We thus conclude that \texttt{NNAD} is particularly suited
for large-scale problems.

We would finally like to stress that we are aware of the plethora of
available minimisers dedicated to problems involving neural networks,
such as \texttt{Tensorflow}~\cite{tensorflow2015-whitepaper},
\texttt{PyTorch}~\cite{NEURIPS2019_9015}, and
\texttt{Keras}~\cite{chollet2015keras}. However, a typical feature of
these tools is that their mainstream user interface is written in
\texttt{python}. Therefore, a direct comparison of our
back-propagation formula to the differentiation strategies of these
codes, would require a \texttt{python} implementation of
\texttt{NNAD}. We do plan to carry out such an implementation, along
with comparisons to the main available tools, in a future publication.

\section*{Acknowledgements}

We thank Nathan Hartland for initiating the idea and collaboration in
the initial stages of this work. We also thank Juan Rojo and Jacob
Ethier for their feedbacks. We would also like to thank the members of
the \texttt{xFitter} Developers' team~\cite{Bertone:2017tig},
particularly Sasha Glazov, for introducing us to
\texttt{ceres-solver}. R.A.K was supported by the Netherlands
Organization for Scientific Research (NWO). V.B. was supported by the
European Research Council (ERC) under the European Union's Horizon
2020 research and innovation program (grant agreement No. 647981,
3DSPIN). This project has received funding from the European Union’s
Horizon 2020 research and innovation programme under grant agreement
No. 824093.

\appendix
\section{General overview of the {\tt NNAD} code}\label{app:code}

In this appendix we discuss the main functionalities of the
\texttt{NNAD} implementation of a feed-forward NN and its analytic
derivatives. An example code can be found in the public distribution
of the \texttt{NNAD} code under the name
\texttt{tests/TestDerivatives.cc}.  The purpose of that code is to
build a NN, compute the NN itself and all its derivatives at some
particular point, and finally compare the results to a fully numerical
calculation.

The first step to take is the definition of the architecture of the NN
in terms of a \texttt{std::vector} of \texttt{int}'s, for example:
\begin{verbatim}
  // Define architecture
  const std::vector<int> arch{3, 5, 5, 3};
\end{verbatim}
This defines the architecture $[3,\,5,\,5,\,3]$ with three input
nodes, two fully-connected hidden layers with five nodes each, and
three output nodes. The next step is to build the NN itself and this
is easily done as follows:
\begin{verbatim}
  // Initialise NN
  const nnad::FeedForwardNN<double> nn{arch, 0, true};
\end{verbatim}
The NN is an object of the template namespaced class
\texttt{FeedForwardNN} that takes as (mandatory) inputs the
architecture and a random seed (\texttt{0} in this case) required to
randomly initialise weights and biases between zero and one. There are
further optional inputs.  The first is a switch to turn on or off the
report of the NN features. This parameter is off by default but it is
turned on in the example above. As a further input, the user can
specify the functional form of the activation function $\phi$ (that is
used for all non-input nodes) and its derivative $\phi'$. It is
responsibility of the user to make sure that activation function and
its derivative correctly match. Both these inputs should be passed as
\texttt{std::function}'s taking a \texttt{double} as an input and
returning a \texttt{double}. The default for these parameters is the
sigmoid function and its derivative:
\begin{equation}
\phi(x)=\frac{1}{1+e^{-x}}\,,\qquad \phi'(x) = \phi(x)\left[1-\phi(x)\right]\,.
\end{equation}
The final optional input is whether the output nodes should use the
same activation function of the hidden nodes or if their activation
function should be linear. The code uses linear activation functions
as a default for the output nodes.

Once the NN object has been built, one needs to define an input that
has to be a \texttt{std::vector} of \texttt{double}'s with as many
elements as input nodes: three, in this case. For example:
\begin{verbatim}
  // Input vector
  std::vector<double> x{0.1, 2.3, 4.5};
\end{verbatim}
There are two main methods to evaluate the NN at \texttt{x}. The first
is:
\begin{verbatim}
  // Get NN at x
  std::vector<double> nnx = nn.Evaluate(x);
\end{verbatim}
that returns a \texttt{std::vector} of \texttt{double}'s with as many
elements as output nodes (three, in this case) corresponding to the NN
values. In order to also access the derivatives, one can call:
\begin{verbatim}
  // Get NN and its derivatives at x
  std::vector<double> dnnx = nn.Derive(x);
\end{verbatim}
Also this function returns a \texttt{std::vector} of \texttt{double}'s
but with as many elements as output nodes plus the number of
parameters times the number output nodes. In fact, the function
\texttt{Derive}, not only returns the derivatives, but also the values
of the NN itself that are placed in the first (three, in this case)
entries of the vector \texttt{dnnx}.

The class \texttt{FeedForwardNN} provides more methods to handle its
objects. A more detailed documentation can be found in the online
repository.

\bibliographystyle{elsarticle-num}
\bibliography{CPC_format}

\end{document}